\documentclass[11pt]{article}
\usepackage[dvips]{color}                 
\usepackage{graphicx}                     
\usepackage{amssymb}
\usepackage{amsmath}
\usepackage{xspace}                       
\newcommand{\vs}{\vspace{-0.25cm}}
\renewcommand{\today}{\ifcase\day\or 1st\or 2nd\or 3rd\or 4th\or 5th\or 6th\or
 7th\or 8th\or 9th\or 10th\or 11th\or 12th\or 13th\or 14th\or 15th\or 
 16th\or 17th\or 18th\or 19th\or 20th\or 21st\or 22nd\or 23rd\or 24th\or
 25th\or 26th\or 27th\or 28th\or 29th\or 30th\or 
 31st\fi~\ifcase\month\or January\or February\or March\or April\or
 May\or June\or July\or August\or September\or October\or November\or
 December\fi \space \number\year}   
\newcommand{\mytitle}[1]{
                         \begin{center}
                           \LARGE{\textbf{#1}}
                         \end{center}}
\newcommand{\myauthor}[1]{\textbf{#1}}
\newcommand{\myaddress}[1]{\textit{#1}}

\begin{document}

\begin{titlepage}

\vspace*{0.5cm}
\mytitle{Cutoff schemes in chiral perturbation  
theory and the quark mass expansion of the nucleon 
mass\footnote{Work supported in part by BMBF and DFG}}
  \vspace*{0.3cm}

\begin{center}
 \myauthor{V{\'e}ronique Bernard$^a$},
  \myauthor{Thomas R. Hemmert$^{b}$}, 
  and
  \myauthor{Ulf-G. Mei{\ss}ner$^{c}$} 

  \vspace*{0.5cm}
\myaddress{$^a$ Laboratoire de Physique Th\'eorique, Universit\'e Louis Pasteur\\
                3-5, rue de l'Universit\'e, F-67084 Strasbourg, France\\
 (Email: bernard@lpt6.u-strasbg.fr)}\\[2ex]
  \myaddress{$^b$
    Physik-Department, Theoretische Physik T39  \\
    TU M{\"u}nchen, D-85747 Garching, Germany\\
    (Email: themmert@physik.tu-muenchen.de)}\\[2ex]
  \myaddress{$^c$ Helmholtz-Institut f\"ur Strahlen- und Kernphysik (Theorie), 
                  Universit\"at  Bonn\\
                  Nu{\ss}allee 14-16, D-53115 Bonn, Germany  \\
    (Email: meissner@itkp.uni-bonn.de)}

  \vspace*{0.2cm}
\end{center}

\vspace*{0.5cm}

\begin{abstract}
\noindent
We discuss the use of cutoff methods in chiral perturbation theory.
We develop a cutoff scheme based on the operator structure of the effective
field theory that allows to suppress high momentum contributions
in Goldstone boson loop integrals and by construction is free of the problems traditional
cutoff schemes have with gauge invariance or chiral symmetries. 
As an example, we discuss the chiral
expansion of the nucleon mass. Contrary to other claims in the literature 
we show that the mass of a nucleon in heavy baryon chiral perturbation theory 
has a well behaved chiral expansion up to effective Goldstone boson masses of 
400 MeV when one utilizes standard dimensional regularization techniques. 
With the help of the here developed cutoff scheme we can demonstrate 
a well-behaved chiral expansion for the nucleon mass up to 600 MeV 
of effective Goldstone Boson masses. We also discuss in detail the prize, in numbers 
of additional short distance operators involved, that has to be paid for this extended 
range of applicability of chiral perturbation theory with cutoff regularization, 
which is usually not paid attention to. We also compare the fourth order result
for the chiral expansion of the nucleon mass with lattice results and draw some
conclusions about chiral extrapolations based on such type of representation.
\end{abstract}
\end{titlepage}
\section{Introduction}
\setcounter{footnote}{0}

Systematic methods of effective field theory (EFT) have established themselves 
over the past 20 years 
in the field of hadronic physics (and many other branches of physics). 
In hadron physics they are based on our understanding that the chiral
symmetry of QCD is spontaneously broken at low energies, leading to the emergence of 
Goldstone bosons. In two--flavor QCD one expects three Goldstone boson modes which 
are identified 
with the physical pions. There exists a mass gap between the pions and the lowest 
lying SU(2) matter fields like the $\rho$--meson or the nucleon. Thus it has been realized 
already a long time ago \cite{Weinberg,GL1,GL2} that a regime of low--energy QCD 
should exist where the dynamics
is governed by the Goldstone boson modes. In that regime an effective field theory of QCD,
usually
called chiral perturbation theory (CHPT), can be set up, which is written in terms of the 
very Goldstone degrees of freedom, with their couplings to matter fields dictated by
chiral dynamics \cite{Wein2,CCWZ1,CCWZ2}.  Recently doubts have been issued in the literature
\cite{Amherst1,Amherst2,Adelaide,Adelaide2}  whether such an effective field theory, in particular 
when it is utilized in connection with 
dimensional regularization, is ``effective'' 
enough to be applied to extended objects with complicated 
internal structure like baryons. The authors of Refs.~\cite{Adelaide,Adelaide2} therefore 
favored the use of a cutoff regularization scheme in their recent analysis of the 
nucleon mass, which in their eyes is more suitable for such a scenario. 
While the use of cutoff methods in CHPT already has a long history 
\cite{GZ,Juerg,Amherst1} and constitutes a well-defined regularization
procedure, from the viewpoint of field theory it is not acceptable 
that one scheme should provide superior results over the other (for a general
discussion of regularization schemes in CHPT, see e.g. \cite{EM}).  

\noindent
In this paper we first want to remind the reader how the internal 
structure of baryons manifests itself in  chiral perturbation theory 
(or any variant thereof) and in addition we will give a detailed 
comparison between dimensional and cutoff regularization for 
a {\em very} simple leading-one-loop order calculation, namely the nucleon mass. 
We proceed to demonstrate the obvious, one-to-one agreement between the cutoff and 
the dimensional regularization result. We then proceed to propose a
novel scheme for cutoff regularization, which allows one to find cutoff--independent 
plateaus over a larger region in the low--energy domain. This scheme is not 
based on any model--dependent framework (like e.g. the implementation of some form factor 
related to the scale set by the hadron size) but rather utilizes directly the 
structure of the effective Lagrangian underlying the effective field theory (EFT) 
under consideration. 
We consider this a distinct advantage over the schemes proposed so far in the literature 
in the context of chiral perturbation theory . However, the involved suppression of 
momentum modes above the scale of validity of the effective field theory 
comes at the cost of additional short distance operators to be taken into account. 
We also extend these considerations to the terms proportional to the quark masses
squared in the chiral expansion of the nucleon mass and discuss the related issue
of chiral extrapolation for lattice results obtained for large pion (quark) masses.

\medskip\noindent
The manuscript is organized as follows. Section~\ref{sec:internal} contains a 
brief discussion of
how the hadron structure emerges in chiral perturbation theory.  Readers familiar with 
the basic concepts of effective field theory may directly proceed to Section~\ref{sec:cutoff}.
There, we give a  comparison between dimensional and cutoff regularization for one particular
observable (the nucleon mass). We also show how the cutoff regulated calculation can be improved
systematically. The chiral expansion of the nucleon mass to fourth
order and the consistency of the resulting chiral expansion with data from lattice gauge 
theory are displayed and
discussed in Section~\ref{sec:mass4}. 
In Section~\ref{sec:test} we then list the criteria how the prediction for any observable
can be made essentially cutoff--independent and how larger plateaus can be obtained. 
A summary and conclusions are given in 
Section~\ref{sec:sum}. The isovector anomalous magnetic moment 
of the nucleon is discussed in the appendix.

\vfill\pagebreak
\section{Reminder: Internal structure of baryons in chiral EFTs}
\label{sec:internal} 

For the study of nucleon structure in chiral effective field theories a Lagrangian formalism is 
set up 
written in terms of spin-1/2 matter fields (nucleons) or sometimes even explicit spin-3/2 states 
($\Delta$(1232)). These matter fields are chirally coupled to the Goldstone
modes and external sources, in harmony with gauge invariance and other symmetries.
Most of the calculations of the past 10 years have 
been performed in a non-relativistic version of such chiral effective field theories, often denoted heavy baryon chiral 
perturbation theory (HBCHPT) or small scale expansion (SSE) (if heavy  spin-3/2
degrees of freedom are explicitely included)\footnote{Recently new
  calculations in a relativistic framework called 
infrared regularization \cite{ET}\cite{BL} (for an extension to spin-3/2
fields, see \cite{BHM})  have 
gained renewed popularity due to poor convergence properties of the 
non-relativistic frameworks in {\em some}  
calculations, in particular the ones regarding the 
low--energy spin structure of the nucleon \cite{spin1,spin2}.} 
\cite{HBCHPT} \cite{SSE}. While it
is true, as sometimes emphasized in the literature, 
that the (bare) baryon fields occurring in the chiral 
Lagrangians correspond to {\em structureless} spin-1/2 or spin-3/2 states 
that at first sight have 
little to do with 
with the complicated hadronic objects we know as nucleons and resonances, this observation 
only holds at tree level utilizing the lowest order chiral and gauge invariant
couplings. 
As soon as one goes beyond the leading order Lagrangians which describe 
physics at the tree level, these objects begin to build up an 
internal structure according to rules dictated 
by chiral symmetry. In general the internal structure of 
baryons manifests itself in two different ways in 
these chiral effective field theories:

\noindent
First, the effective Lagrangian contains all terms allowed by 
PCT-transformations and chiral symmetry. 
This leads to a string of terms of higher order couplings which usually are 
termed ``counterterms'' in 
the literature independent of their respective role in the renormalization procedure. 
These couplings
accompany higher dimensional operators and parameterize all short distance 
physics contributing to the internal structure of the baryons. 
We note that these higher order operators related 
to the internal structure of the nucleon, the low--energy constants (LECs), already 
occur at next-to-leading order (NLO) in the chiral Lagrangian for baryons
(these LECs are usually denoted as $c_i$  \cite{HBCHPT}), whereas the famous chiral 
loops only start to contribute at N$^2$LO and are connected to a different set of 
LECs denoted $B_i$ in the earlier literature and nowadays often $d_i$. 
For example, the finite size of the "core" of a nucleon which 
plays a big role in the (chiral) bag model would have its analogue in the term proportional
to the third order LEC $B_{10}$  of the isovector 
Dirac radius of the nucleon discussed below, cf. Eq.(\ref{radius}) given below. 
The effective field theory therefore contains all the 
required terms to account for a non-pointlike baryon, which should not 
come as a surprise because it 
has to represent a complete mapping of the underlying QCD Lagrangian 
responsible for the intricate 
structure of baryons into the low--energy domain. 

\noindent
A second finite size effect of the nucleon only starts to get generated at the 
one-loop level, as  in every quantum field theory short--lived fluctuations can occur. 
In baryon CHPT a nucleon typically 
emits a pion, this energetically forbidden $\pi$N intermediate state lives for a 
short while and then 
the pion is reabsorbed by the nucleon, in accordance with the uncertainty principle. 
This mechanism is 
responsible for the venerable old idea of the ``pion cloud'' 
of the nucleon, which in CHPT can be put on 
the firm ground of field theoretical principles. It is clear that this effect 
generates the longest range 
finite size effect connected to the internal structure of baryons due to the 
lightness of the pion as  the (nearly massless) Goldstone boson. Furthermore, 
{\em only} such pion loop effects can lead to
terms that are non--analytic in the quark masses, famous examples are the 
baryon mass terms $\sim
m_\pi^3$, the nucleon isovector Dirac radius $ \sim \ln m_\pi$ or the nucleon´s 
electromagnetic 
polarizabilities $ \sim 1/ m_\pi$. In the last two cases, these contributions even  
become singular 
in the chiral limit $m_\pi \to 0$ and dominate the corresponding n--point
functions. 
To be more specific,
we note that calculations of the  nucleon radii in CHPT show that a sizeable part of 
the nucleon size, as for example seen in 
electron scattering, originates from the pion cloud (at least in the isovector channel). 
{\em However}, 
such a statement of course always depends on the value of the regularization scale used. 
For a concrete  example we discuss the isovector Dirac radius of the nucleon to leading
one-loop order in HBCHPT \cite{BKKM}: 
\begin{eqnarray}
\left(r_1^v\right)^2 &=& -\frac{1}{(4\pi F_\pi)^2}\left\{1+7g_A^2+\left(10g_A^2+2\right)
\log\left[\frac{m_\pi}{\lambda}\right]\right\}-12 B_{10}(\lambda)\nonumber\\
 &=& \left(0.61-0.47\,\rm{GeV}^{-2} \, B_{10}(\lambda)+0.47 \log\frac{\lambda}{1\rm{GeV}}
\right)\rm{fm}^2
\label{radius}
\end{eqnarray}
Here, $g_A=1.267$ denotes the axial--vector coupling of the nucleon determined from 
neutron beta decay, 
$F_\pi=92.4$ MeV is the (weak) pion decay constant and $m_\pi=138$ MeV corresponds 
to the pion mass. 
Furthermore, $\lambda$ is the scale of dimensional regularization. As 
discussed in the previous paragraph, the LEC $B_{10}(\lambda)$ parameterizes any short 
distance contributions to 
the Dirac radius, resolved at the regularization scale $\lambda$. 
Compared to the empirical value 
$(r_1^v)^2=0.585$ fm$^2$ \cite{MMD,HMD} we note that several combinations 
of $(\lambda,B_{10}(\lambda)$  pairs can reproduce the empirical result, e.g.
\begin{eqnarray}
\left(1~\rm{GeV},+0.06~\rm{GeV}^{-2}\right),\left(0.943~\rm{GeV},\rm{0.00~GeV}^{-2}\right),
\left(0.6~\rm{GeV},-0.46~\rm{GeV}^{-2}\right)\, .\label{pairs}
\end{eqnarray}
An important observation to make is that even the sign of the ``core'' contribution 
to the radius 
described by $B_{10}$ can change within a reasonable range typically used for $\lambda$. The 
question of whether explicit $\Delta$(1232) degrees of freedom have to be added in this 
calculation or 
not does not change the situation at all and is irrelevant from the point of field 
theory\footnote{For corresponding $(\lambda,B_{10}(\lambda)$ pairs in a theory with 
explicit $\Delta$ 
degrees of freedom see Ref.~\cite{BFHM}.}. Physical intuition would tell us that the 
value for coupling $B_{10}$ 
should be negative such that the nucleon core gives a {\em positive} contribution to 
the isovector 
Dirac radius, see Eq.(\ref{radius}), but field theory tells us that for (quite reasonable) 
regularization scales above $\lambda=943$ MeV this need not be the case. In general it is our 
observation that if one really wants to make {\em qualitative} contact with phenomenological 
models one  should use rather small values $\lambda\approx 600$ MeV in dimensional 
regularization. However, from the point of 
view of field theory all these choices of Eq.(\ref{pairs}) are of course equivalent 
and all three of them describe the same 
physics and {\em all of them contain the same amount of information about the 
finite size structure of 
the nucleon,} independent of our ability or personal preference for interpreting them. 
We emphasize  again that only the sum of $B_{10}(\lambda)$ and the associated scale--dependent 
chiral logarithm constitutes a meaningful quantity that can be discussed. 

\begin{figure}[tb]
  \begin{center}
    \includegraphics*[width=0.9\textwidth]{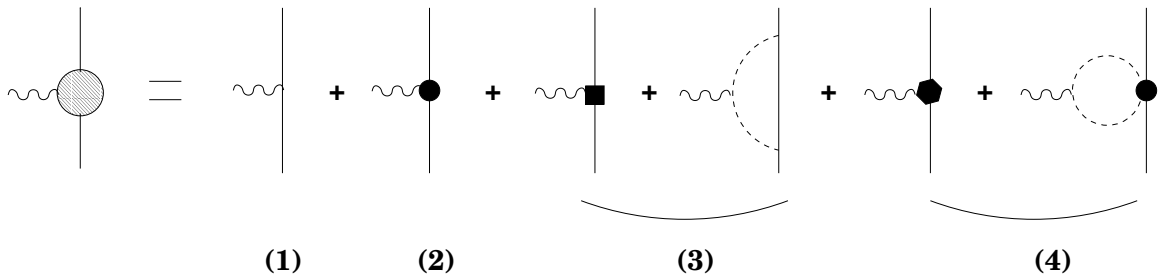}
    \caption{Chiral expansion of a baryon electromagnetic form factor. The
      lowest order graph (1) with dimension one insertions only gives the charge
      of the baryon. Tree graphs with insertions from the dimension two (2), three (3)
      and four (4) effective Lagrangian are depicted by the solid circle, square
      and sextangle, in order. Pion loop graphs start at third order (3), and
      fourth order loop graphs (4) have exactly one dimension two insertion. Only
      one typical loop graph at the orders considered is shown. Solid,
      dashed and wiggly lines denote baryons, pions and photons,
      respectively.}
\label{ff}
  \end{center}
\vspace{-0.3cm}
\end{figure}

\medskip\noindent
We can summarize these remarks in the cartoon of any nucleon electromagnetic
form factor shown in Fig.~\ref{ff}. It consists of local operators with
insertions from the chiral effective Lagrangian with increasing dimension and
the pion cloud contribution, which starts at third order and is given by  
another string of terms with increasing dimension. In our example, the nucleon
charge is given by the lowest order electric coupling corresponding to
graph~(1) in  Fig.~\ref{ff} whereas the magnetic couplings only start at second
order corresponding to the graphs~(2). Further contributions to the magnetic 
moments are given by pion loop graphs at third (and higher) order. At third (fourth)
order, we also have the first contributions to the electric (magnetic) nucleon
radii. Higher curvature terms are only build up at higher orders, see e.g. the
discussion in \cite{KM}.

\bigskip\noindent
For practitioners in chiral EFTs the points discussed in this section 
are common knowledge. In fact, the vast 
majority of processes calculated in HBCHPT at the one-loop level (e.g. $\pi N\to\pi N,\;\pi 
N \rightarrow\pi\pi N,\; \gamma N\rightarrow\pi N,\;\gamma^\ast N\to \pi N,\ldots$) behave 
completely analogous to the scenario
displayed in Eq.(\ref{radius}) for the isovector Dirac radius. 
In such type of scenarios dimensional 
regularization is the effective method of choice, all symmetries of the theory are 
automatically preserved and the entire information (up to a given order in the perturbative 
calculation) on short and long distance physics is contained in the result, independent 
of the particular regularization  scale chosen. However, there are some one-loop calculations 
that at first sight appear to generate a 
scale--independent result. The question arises how the discussion given here translates 
to these special  cases where only polynomial divergences appear in the calculation 
of the loop integrals. These  polynomial divergences are automatically eaten up in 
dimensional regularization, providing us with a nice, seemingly scale-free one--loop 
prediction. The most prominent example, the leading-one-loop 
pion contribution to the mass of the nucleon, is discussed in the next section. 

\section{A comparison of dimensional and cutoff regularization}
\label{sec:cutoff}

\subsection{Third order calculation of the nucleon mass in dimensional regularization}

\begin{figure}[tb]
  \begin{center}
    \includegraphics*[width=0.5\textwidth]{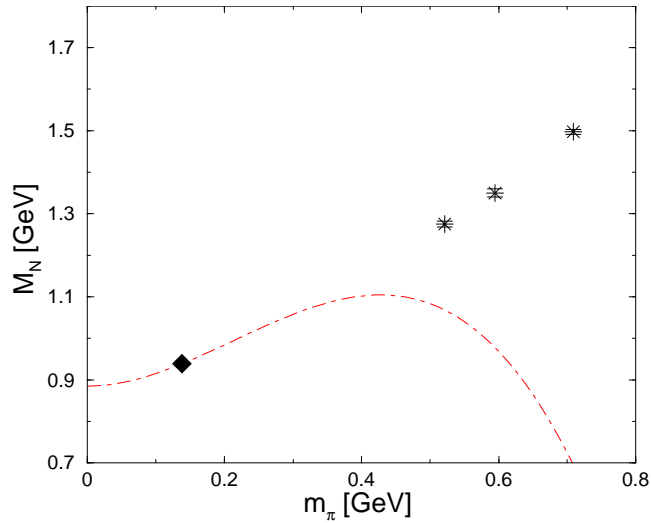}
    \caption{Mass of the nucleon calculated to ${\cal O}(p^3)$ in HBCHPT plotted 
as a function of the mass of the pion. The parameter choices made are discussed in the 
text. One can clearly see that this simple leading one-loop calculation breaks down 
around $m_\pi=400$ MeV. The lattice data (stars) shown have been selected from 
Ref.~\protect\cite{Adelaide2,CPPACS}. The filled diamond gives the nucleon mass at the
physical value of the pion mass.}
\label{p3plot}
  \end{center}
\vspace{-0.3cm}
\end{figure}

The leading one--loop calculation of the mass of the nucleon in HBCHPT gives 
\begin{eqnarray}
M_N^{(3)}&=&M_0-4\,c_1 m_\pi^2-\frac{3\,g_A^2}{32\pi F_\pi^2}\,m_\pi^3+{\cal O}(p^4)\;.
\label{HBmass}
\end{eqnarray}
Here, $p$ denotes a small parameter (external momentum, meson mass) and
$M_0$ corresponds to the nucleon mass in the chiral limit. 
Its precise value is not well known. 
Various analyses \cite{masses1,masses2,masses3,masses4} 
suggest values in the range $0.75\ldots0.9$ GeV.  It is hoped that new 
lattice analysis will provide a better understanding of this 
important nucleon structure parameter  \cite{Gerrit}. 
Similarly, $g_A$ and $F_\pi$ should be taken at their values in the chiral limit, but 
for our arguments this fine point is of no relevance (as the differences only show
up at higher orders).
$c_1$ denotes a dimension two coupling in the chiral Lagrangian sensitive to the internal 
structure of the nucleon in close connection to the pion--nucleon sigma term. 
Most determinations agree about  its sign but are uncertain about its precise value
(related to the still unsatisfactory situation in the extraction of the pion-nucleon 
sigma term based on various partial waves analyses of elastic $\pi $N 
scattering data): $c_1=-0.9\pm 0.5$ GeV$^{-1}$ \cite{c11,c12,c13,c14,c15,c16,c17}. 
Choosing the central value for $c_1$ and fixing $M_0=0.88$ GeV 
to reproduce the physical mass of the nucleon for 
$m_\pi=0.138$ GeV one therefore obtains the mass of the nucleon as a function of the 
mass of the pion as shown in 
Fig.~\ref{p3plot}. Since the result for the one-loop mass shift is finite, 
the only dimensionfull
scale in the problem is the pion mass. The term proportional to a fractional
power of the quark masses is generic to the chiral expansion of the octet 
baryon masses in terms
of pion, eta and kaon Goldstone bosons. In lattice gauge studies one
usually has to deal with rather large pion masses and thus is interested in 
a theoretically well-founded extrapolation function to obtain a prediction for 
an observable at the physical pion mass.
From Fig.~\ref{p3plot} one clearly sees that  Eq.~(\ref{HBmass}) is 
only able to give a sensible description of the nucleon mass may be out to pion masses of 
400 MeV. In the next section we will demonstrate that this breakdown of the 
leading one-loop HBCHPT result for Goldstone boson masses in this range is connected 
with the rise of short distance physics contributions, which are not properly controlled 
in standard dimensional regularization for $m_\pi>400$ MeV. 
Without giving any details we just note that the addition of explicit 
$\Delta$(1232) degrees of freedom in the non-relativistic SSE scheme does not 
dramatically improve this situation.

\noindent
While Fig.~\ref{p3plot}  might be a disappointment to any QCD lattice practitioner 
who would  like to extrapolate large quark mass simulations of the nucleon mass 
down to the physical point utilizing Eq.(\ref{HBmass}), from the point of view of 
CHPT such a behaviour is entirely consistent and in some sense even expected. 
It would be entirely unreasonable to assume that this leading non-analytic pion mass 
dependence calculated at the leading-one-loop level in HBCHPT should 
be enough to describe quark-mass dependent physics far above the physical pion 
mass, as witnessed by the fact that there are large fourth order corrections
to the baryon masses. \footnote{Another
example is given by the  recent studies of chiral extrapolation functions 
of the isovector anomalous magnetic moment of the nucleon and of the axial coupling 
of the nucleon, where it was argued that the respective 
leading one-loop ${\cal O}(p^3)$ HBCHPT results calculated to the same accuracy 
as the mass formula of 
Eq.(\ref{HBmass}) do not contain enough quark-mass dependent structures to 
connect between the chiral limit values of these quantities to the physical point,
assuming that one can apply such one--loop formulae at very large pion masses as done
there \cite{extrapolation1,extrapolation2}.} 
In the present context we now want to study the question whether the same
leading one--loop 
${\cal O}(p^3)$ calculation of the same diagram miraculously can give a much better behaved 
result for 
large pion masses when the calculation is performed with the help of a cutoff 
regularization procedure 
instead of the so-far employed dimensional regularization. This might at first
sound strange to any practitioner of field theory, but in the context of an
underlying power counting a different regularization scheme might lead to a
reordering of the expansion and thus improve the convergence. 

\subsection{Third order calculation of the nucleon mass in cutoff regularization}

Evaluating the same diagrams as before with the help of a 
radial cutoff $\Lambda$ one obtains the following finite expression in 
HBCHPT \footnote{We are utilizing a non-covariant cutoff approach which 
only acts in momentum space. Obviously many different versions of cutoff implementations, e.g. 
covariant ones, exist in the literature. The points made here apply across 
all these approaches.}:
\begin{eqnarray}
M_N^{(3)}&=&M_0^{(r)}-4c_1^{(r)}\,m_\pi^2-\frac{3\,g_A^2}{32\pi F_\pi^2}\,m_\pi^3 
                +\frac{3\,g_A^2}{(4\pi F_\pi)^2}\,m_\pi^3
                \arctan\frac{m_\pi}{
                \Lambda}\label{HBlambda}
\end{eqnarray}
with
\begin{eqnarray}\label{renorm}
M_0^{(r)}&=&M_0-\frac{g_A^2}{(4\pi F_\pi)^2}\,\Lambda^3\, , \nonumber\\
c_1^{(r)}&=& c_1 - \frac{3\,g_A^2}{4 (4\pi F_\pi)^2}\,\Lambda\, .
\end{eqnarray}
Having absorbed the polynomial  divergences in 
$\Lambda$ into the coefficients $M_0$ and $c_1$, we have regulated the high-energy 
behaviour and obtained a finite result. Note that in dimensional regularization, 
the parameters at dimension one and two of the effective Lagrangian do not get renormalized.
However, one should keep in mind that the effective field theory is 
only defined for momenta $p \leq\Lambda_\chi\approx 1$ GeV. For 
$\Lambda \geq \Lambda_\chi$ 
we must therefore be able to map the residual cutoff dependence onto a 
string of higher order couplings that are suppressed by inverse powers of $\Lambda_\chi$. 
One finds

\begin{figure}[t]
  \begin{center}
    \includegraphics*[width=0.5\textwidth]{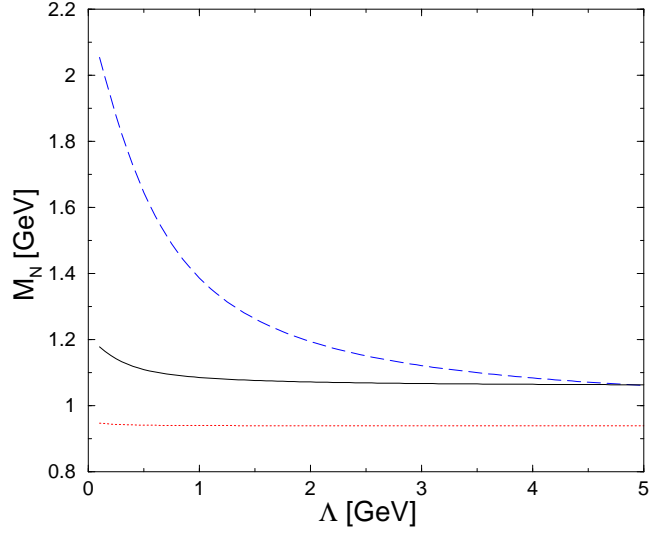}
    \caption{Cutoff dependence of the nucleon mass calculated at ${\mathcal O}(p^3)$ in 
             HBCHPT. The dotted, solid and dashed curve correspond to a pion mass of 140, 
             300 and 600 MeV, in
             order. No plateau is found in the region of applicability of CHPT 
             (i.e. for $\Lambda <1$\,GeV) for $m_\pi > 400\,$MeV.}
\label{dependence}
  \end{center}
\end{figure}

\begin{eqnarray}
M_N^{(3)}&=&M_0^{(r)}-4\,c_1^{(r)}\,m_\pi^2-\frac{3\,g_A^2}{32\pi F_\pi^2}\,m_\pi^3 
                +\frac{3\,g_A^2}{(4\pi F_\pi)^2}\,m_\pi^3\left\{\frac{
                m_\pi}{\Lambda_\chi}-\frac{1}{3}\,\left(\frac{m_\pi}{
                \Lambda_\chi}\right)^3+\ldots\right\}\nonumber\\
             &=&M_0^{(r)}-4\,c_1^{(r)}m_\pi^2-\frac{3\,g_A^2}{32\pi 
                F_\pi^2}\,m_\pi^3
                +{\cal O}(p^4)\;.\label{HBcutoff}
\end{eqnarray} 
To third order ${\cal O}(p^3)$ one therefore obtains {\em exactly} the same 
result\footnote{Keeping the full 
$\arctan$ function (Eq.~(\ref{HBlambda})) instead of systematically truncating 
at ${\cal O}(p^3)$ 
(Eq.~(\ref{HBcutoff})) is equivalent to modeling some $p^4,\, p^5,\ldots$ 
contributions, as this 
procedure for example cannot account for all $p^4$ structures generated via a 
systematic calculation.} 
in cutoff and in dimensional regularization, as expected (for a related discussion
between dimensionally and cutoff regularized loop integrals in chiral nuclear EFT, see
\cite{EGMreg}). We note that the terms beyond the 
${\cal O}(p^3)$ truncation only contain even powers of $m_\pi$, showing that 
the cutoff scheme employed 
here respects chiral symmetry at the one--loop level, since such terms are 
analytic in the quark masses and can thus be
absorbed in the polynomial terms generated by the effective Lagrangian. 
We note that most of the regulating functions for one-loop diagrams proposed in 
\cite{Adelaide} do not share this property, but generate e.g. terms 
$\sim m_\pi^5/\Lambda_\chi^2$, which can only be mapped onto a systematic chiral 
expansion at the two-loop level. In conclusion we note that formally the correspondence
between the results obtained in dimensional and cutoff regularization 
is obtained by an expansion of the cutoff result in powers of $1/\Lambda=1/\Lambda_\chi$.  

\noindent
In Fig.~\ref{dependence} we now show the resulting cutoff dependence of the 
nucleon mass to leading one-loop order in HBCHPT of Eq.(\ref{HBlambda}) for various values
of the pion mass. As expected, for the physical value of $m_\pi$, we have a nice 
plateau for all cutoff
values below $\Lambda_\chi$, clearly showing that CHPT for two flavors is well behaved. 
For $m_\pi = 300$ one observes already some curvature in the range of reasonable cutoffs, 
whereas for a pion mass as large as 600~MeV, there is no stable prediction in the range
of physically allowed cutoffs. We therefore observe that the breakdown of the leading
one-loop HBCHPT result of dimensional regularization as seen in Fig.~\ref{p3plot} 
and the disappearance of a stable plateau in the low energy region in the cutoff approach,
cf. Fig.~\ref{dependence}, occur in the same mass range. Both phenomena point to the fact 
that for Goldstone Boson masses larger than 400 MeV the standard counting of CHPT 
has be modified to properly account for the short distance physics. 
In the next section we will present such a modification within the cutoff approach.

\subsection{Introduction of an improvement term}
\label{sec:improved}

\begin{figure}[t]
  \begin{center}
    \includegraphics*[width=0.5\textwidth]{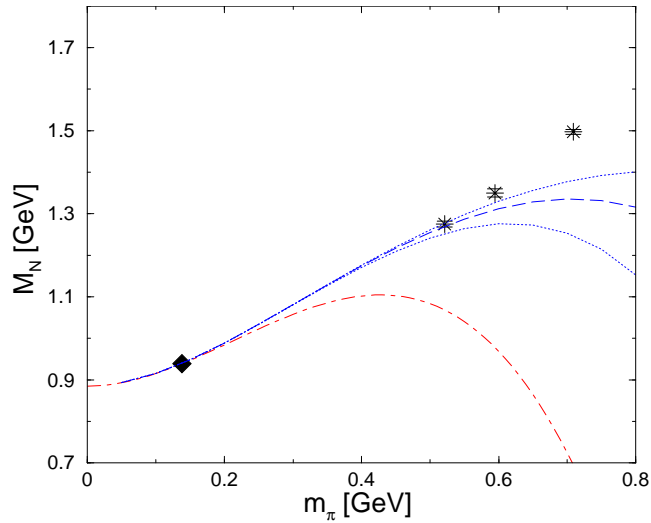}
    \caption{Nucleon mass calculated to third order in HBCHPT as a function of the pion
             mass. The solid curve shows the ${\mathcal O}(\Lambda^{-1}$) improved scheme,
             whereas the dotted curve shows the result in standard
             dimensional regularization. The solid line refers to a cutoff $\Lambda = 1\,$GeV,
             whereas the dotted lines give the corresponding results for $\Lambda = 0.8\,$GeV
             and $\Lambda = 1.2\,$GeV, respectively.
             For further notation see Fig.~\ref{p3plot} }
\label{mpiplot}
  \end{center}
\end{figure}

The expansion of the nucleon mass using cutoff regularization given in  
Eq.~(\ref{HBcutoff}) allows us
to propose an improvement scheme to avoid the strong cutoff dependence for larger pion masses
exhibited in Fig.~\ref{dependence}. This can be done entirely within the
operator structure of the EFT under consideration and {\em without} resorting
to phenomenological and model--dependent methods like additional form factors, 
with a cutoff scale possibly inspired by the scale set by the hadron size.
Clearly, the improvement term we are after has to fulfill two requirements:
First, it must lead to the same structure as the leading $1/\Lambda$
correction in  Eq.~(\ref{HBcutoff}) and, second,  must allow us to cancel the strong 
cutoff sensitivity induced by this term. In fact, the corresponding term
appears in the chiral pion--nucleon Lagrangian at fourth order as combination
of three operators. Following the notation of Ref.~\cite{FMMS} these read
\begin{equation}
{\cal L}_{\pi N}^{(4)} = e_{38} \, \bar\psi\,\langle \chi_+ \rangle^2 \, \psi +
e_{115} \, \bar\psi\, \frac{1}{4} \langle  \chi_+^2 - \chi_-^2 \rangle \, \psi -
e_{116} \, \bar\psi\, \frac{1}{4} \biggl(  \langle  \chi_-^2 \rangle -
\langle  \chi_- \rangle^2 +  \langle  \chi_+^2 \rangle - \langle  \chi_+ \rangle^2
\biggr)\, \psi~,
\end{equation}
where $\psi$ denotes the nucleon field and $\chi_\pm$ parametrizes the explicit chiral
symmetry breaking through the quark masses (for precise definitions, see \cite{FMMS}).
These three terms combine to give one operator without pion fields, 
\begin{equation}
{\cal L}_{\pi N}^{(4)} = 4 e_1 \, \bar\psi\, m_\pi^4 \, \psi~,
\end{equation} 
with $e_1 = 16 e_{38} + 2e_{115} + e_{116}/2$.
Furthermore,  the  LEC $e_{1}$ can be written in terms of a finite term and a
second contribution that cancels the leading $1/\Lambda$ correction in
Eq.~(\ref{HBcutoff}), specifically 
\begin{equation}\label{term}
e_{1} = e_{1}^{\rm fin}  - \frac{3 g_A^2}{16 \pi^2 F_\pi^2} \, \frac{1}{\Lambda}~.
\end{equation}
Note the minus sign in front of the second term in this equation, it follows
from the requirement that the leading $1/\Lambda$ correction should be canceled.
Although formally of higher order, one should consider such a term as being promoted to
the order one is working (that is to an order below its formal appearance) 
if one insists on a cutoff--independent result.
\begin{figure}[t]
  \begin{center}
    \includegraphics*[width=0.5\textwidth]{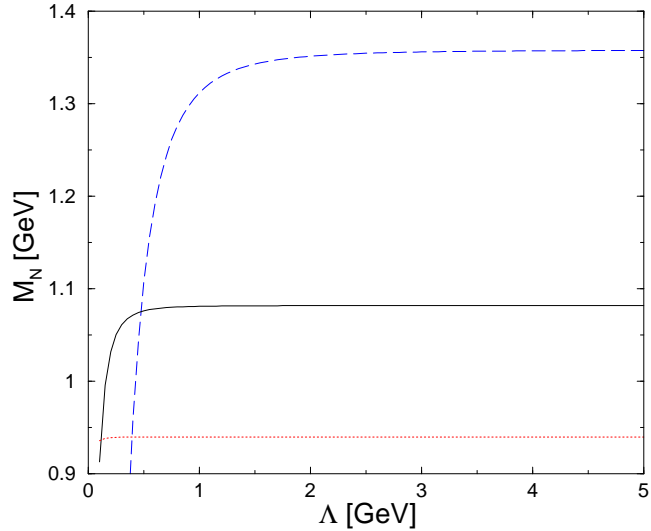}
    \caption{Cutoff dependence of the nucleon mass calculated at ${\mathcal O}(p^3)$ 
             in HBCHPT employing the improvement scheme. The
             dotted, solid and dashed curve correspond to a pion mass of 140, 300 and 
             600 MeV, in
             order. A plateau in the region of applicability of CHPT (i.e. for $\Lambda
             <1$\,GeV) can now be found for Goldstone boson masses up to  $300\,$MeV.}
\label{improved}
  \end{center}
\end{figure}
To show the effect of adding this improvement term, we display in
Fig.~\ref{improved} the cutoff dependence of the nucleon mass calculated to
third order supplemented with the improvement term Eq.~(\ref{term}) and
setting the finite piece to zero, $ e_{1}^{\rm fin} = 0$, for
simplicity. For the physical value of the pion mass, the plateau is, of
course, not affected. However, we now also observe a nice plateau for cutoff
values between 400~MeV and $\Lambda_\chi$ for the larger pion mass $m_\pi =
300\,$MeV. For the even larger pion mass of about 600~MeV, this simple
procedure is still not sufficient to lead to a plateau at sufficiently low
values of the cutoff, but compared to the calculation shown
in Fig.~\ref{dependence} we notice a much less pronounced cutoff sensitivity. 
As a consequence, one can also trust the quark mass expansion of the nucleon
mass to higher pion masses as shown in Fig.~\ref{mpiplot}. Here, we have adjusted
$e_{1}^{\rm fin}$ such as to have the correct mass of the nucleon for the physical
value of the pion mass. It is clear from that figure that the third order contribution
is now reduced compared to the result based on dimensional regularization, however,
one still has to be careful in applying this improved form at too large pion masses,
simply because the term linear in the quark masses becomes very large for pion masses
above 500~MeV.

\noindent
Clearly, this procedure should be considered as the minimal method to improve 
calculations in cutoff schemes. We stress again that all operators required to
achieve such an improvement are part of the effective Lagrangian under consideration
and thus automatically chiral symmetry and the pertinent Ward identities are respected.
It is also important to stress that such an improvement does not come for free -
one has to include one more higher order operator with a tunable finite coefficient,
e.g. in the example just discussed, one would adjust $ e_{1}^{\rm fin}$ so as to 
reproduce the physical nucleon mass for the cutoff under consideration (as it was
done for example to generate Fig.~\ref{mpiplot}). From the
purists point of view, this procedure is of course not entirely satisfactory since
there are other terms at the same order as the improvement term which should 
eventually be considered. However, our ambition is more modest, we simply want to
set up a simple scheme within the EFT to reduce the possible cutoff dependence in
such schemes. Note that similar methods are also used in pionless nuclear
effective field theory approaches to few--nucleon systems, see e.g. \cite{hwh1,hwh2}. 
In the appendix, we briefly discuss the third order result for the nucleon
isovector anomalous magnetic moment employing dimensional and (improved) cutoff
regularization.

\noindent
We now proceed to analyze the nucleon mass to fourth order in the chiral
expansion, that is to second order in the quark masses. It is a well
established fact from many studies of Goldstone boson effects in baryon masses
that such effects can not be neglected, see e.g. \cite{Juerg,masses3}. 
Of course, in these studies, large 
Goldstone boson loops are generated from intermediate kaons and etas whereas
we are considering the case of increasing the pion mass within two--flavor
QCD. Apart from group structure complications, these two cases are equivalent.

\section{Nucleon mass to fourth order and  chiral extrapolation}
\label{sec:mass4}

In this section, we  discuss the nucleon mass shift to fourth order in the chiral
expansion and how it can be used (or misused) to perform chiral extrapolations for
lattice gauge theory results obtained at much larger pion masses than the physical
value. We stress that even the improved one--loop representations discussed here
should not be used at pion masses above $600\ldots  700$~MeV, since one then 
is too close to the
breakdown scale of the chiral expansion. Still, for purely illustrative purpose, we
will apply our formalism to lattice results for pion masses ranging from 500~MeV to
800~MeV.

\subsection{Fourth order calculation of the nucleon mass}
\label{sec:mass4ex}
\begin{figure}[t]
  \begin{center}
    \includegraphics*[width=0.5\textwidth]{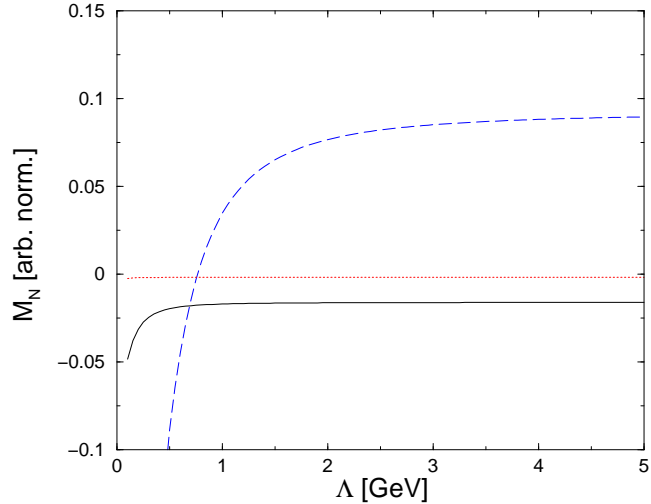}
    \caption{Cutoff dependence of the nucleon mass calculated at ${\mathcal O}(p^4)$ 
             in HBCHPT employing cutoff regularization with $\Lambda = 1\,$GeV
             (in arbitrary normalization).
             For notation, see Fig.~\ref{dependence}.}
\label{figcut4plat}
  \end{center}
\end{figure}

As was derived in \cite{SFM} the fourth order mass shift for a nucleon with
four--velocity $v_\mu$ and residual momentum $k_\mu$ is given 
by:
\begin{equation}
M^{(4)}=i\, \left(\Sigma^{(4)}(\omega=0)-4c_1m_\pi^2 \Sigma'^{(3)}(\omega=0)\right)~,
\end{equation}
where $\omega = v\cdot k$, $\Sigma^{(4)}(\omega )$ is the self--energy to 
fourth order and $\Sigma'^{(3)}(\omega )$
the derivative with respect to $\omega$ of the  self--energy to third order.
In dimensional regularization, the nucleon mass to fourth order takes the form \cite{SFM}
\begin{eqnarray}
M_N^{(4)} &=& M_0-4\,c_1 m_\pi^2-\frac{3\,g_A^2}{32\pi F_\pi^2}\,m_\pi^3
-4\,e_1 (\lambda) m_\pi^4 + \frac{3}{128\pi^2 F_\pi^2}\left( c_2 -
  \frac{2g_A^2}{M_N}\right)\,  m_\pi^4 \nonumber \\
&& \qquad -\frac{3}{32 \pi^2 F_\pi^2} \, \left( -8c_1 + c_2 + 4c_3 + \frac{g_A^2}{M_N}\right)
\, m_\pi^4 \, \ln \frac{m_\pi}{\lambda}
 +{\cal O}(p^5)\;,
\label{HBmass4}
\end{eqnarray}
where  $e_1 (\lambda)$ is a combination of LECs from the fourth order chiral Lagrangian,
as discussed earlier. As shown below, the numerical value of this LEC is not well
known, so for a first orientation we will vary this LEC in the range from $-1$ to
$1$ (in units of GeV$^{-3}$), which are values of natural size, always setting
$\lambda = 1\,$GeV. That, however, is a very conservative estimate and we will
later at least fix the sign of $e_1$. For the present discussion, it is, however,
instructive to work with the full range consistent with naturalness.
The dimension two LECs $c_2, c_3$ are known to some accuracy, the precise values we use will be
given later. Note also that  fifth order corrections to Eq.~(\ref{HBmass4})
have been worked out \cite{MGB} and found to be extremely small for the
physical value of $m_\pi$ and, furthermore, these $1/M_N$ corrections are
given parameter--free.
Shifting now to  cut-off regularization, one obtains for the fourth order
contribution to the nucleon mass
\begin{eqnarray}
\delta M_N^{(4)}&=& -{\frac{3 }{32 \pi^2 M_N F_\pi^2}} \biggl\{
\frac{1}{{3 \sqrt{1+m_\pi^2/\Lambda^2}}} \bigl[ \Lambda^4(
1-\sqrt{1+m_\pi^2/\Lambda^2})(g_A^2 -2M_N(c_2+4c_3)) 
\nonumber \\
& & 
+m_\pi^2 \Lambda^2(g_A^2(1- \frac{1}{2}\sqrt{1+m_\pi^2/\Lambda^2})
+M_N (c_2+4c_3)(1-2\sqrt{1+m_\pi^2/\Lambda^2}) 
\nonumber \\
& &
-24 M_N c_1(1-\sqrt{1+m_\pi^2/\Lambda^2}))
+3m_\pi^4(g_A^2+ M_N(c_2+4c_3-8c_1))\bigr]
\nonumber \\
& &
+(g_A^2-M_N (8c_1-c_2-4c_3)) m_\pi^4\, \biggl[\ln\frac{m_\pi}{M_0} - \ln
\frac{1}{2} \biggl(1  + \sqrt{1+\frac{m_\pi^2}{\Lambda^2}} \biggr)\biggr]
\biggl\} 
- 4 e_1^{(r)}\, m_\pi^4 ~,
\end{eqnarray}
with a further renormalization of the constants $M_0^{(r)}$ and $c_1^{(r)}$ of Eq.~(\ref{renorm}),
\begin{eqnarray}
M_0^{(r)} &=&-{\frac{1 }{32 \pi^2 M_N F_\pi^2}} \Lambda^4
\left( g_A^2 -2 M_N (c_2+4c_3) \right)~,
\nonumber \\
c_1^{(r)} &=&-{\frac{1 }{64 \pi^2 M_N F_\pi^2}}m_\pi^2 \Lambda^2 \left(g_A^2
+ M_N (-48c_1+4c_2+16c_3) \right)~,
\end{eqnarray}
and an additional renormalization of the LEC $e_1$,
\begin{equation}
 e_1^{(r)} = e_1 + \frac{3}{128\pi^2 M_N F_\pi^2}\bigl( g_A^2 - M_N
(8c_1-c_2-4c_3) \bigr) \, \ln \frac{\Lambda}{M_0}~.
\end{equation}
As before, having absorbed the polynomial as well as the logarithmic divergences in 
$\Lambda$ into the coefficients $M_0$, $c_1$ and $e_1$, 
we have regulated the high-energy  behaviour and obtained a finite result.
We note that the fourth order contribution $\delta M^{(4)}$ 
is given by  a polynomial in even powers of $m_\pi$ plus a 
non--analytic piece. This last one is exactly the same as obtained in
dimensional regularization and thus the cut-off scheme preserves chiral
symmetry up to fourth order. 
In contrast to the third order case discussed earlier, 
one observes a very weak cutoff dependence of the
nucleon mass as $m_\pi$ increases as long as $m_\pi \le 400\,$MeV,
as shown in Fig.~\ref{figcut4plat}. If one also wants to obtain a plateau
for larger pion masses below the chiral symmetry breaking scale,
one proceeds in the same way as done before, namely by
adding a six order counterterm such that it has a contribution that
cancels the leading $1/\Lambda^2$ correction:
\begin{equation}
\delta M^{(4)}_{\rm imp}= \delta M^{(4)} + d m_\pi^6~,
\end{equation}
with
\begin{equation}
d=d^{\rm {fin}}- {\frac{5}{512 \pi^2 M_N F_\pi^2 \Lambda^2 }}
\biggl(7g_A^2- 4 M_N \bigl({\frac{36}{5}}c_1 -c_2 -4c_3 \bigr)\biggr)~.
\end{equation}
As before, the so improved result shows a much weaker cut--off dependence.

\subsection{Numerical results and chiral extrapolation}
\label{sec:mass4res}

\begin{figure}[htb]
  \begin{center}
    \includegraphics*[width=0.5\textwidth]{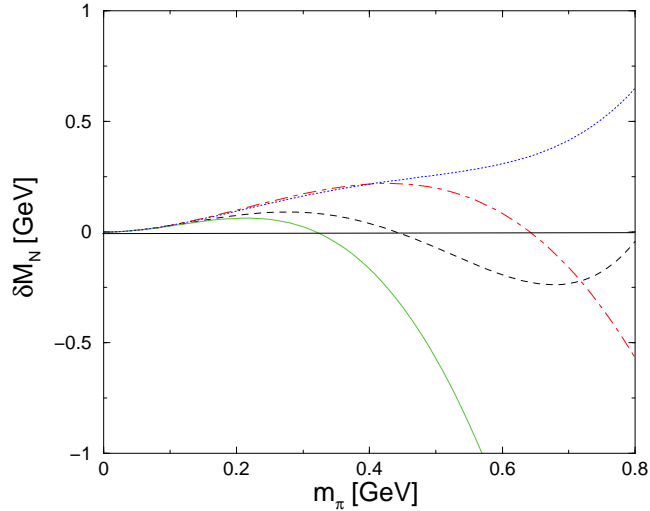}
    \caption{Pion mass dependence of the nucleon mass shift $\delta M_N =
     \delta M_N^{(2)}+ \delta M_N^{(3)}+\delta M_N^{(4)}$. The dot--dashed line
     gives the third order mass shift, whereas the fourth order shift is given
     by the solid and dashed lines for $e_1 = 1$ and $e_1 = -1$, respectively.
     The dotted line gives the fourth order result using $e_1 = -1$ and 
     $c_3 = -4$ (in appropriate units of inverse GeV).}
\label{figmass4}
  \end{center}
\end{figure}

For discussing the numerical results, we have to give the values of the 
LECs appearing in the chiral expansion.
We use here $c_1 = -0.9\,$GeV$^{-1}$, $c_2 = 3.2\,$GeV$^{-1}$ and $c_3 = -5.7\,$GeV$^{-1}$
(and also some smaller (in magnitude) values for $c_3$ 
as discussed below), see e.g. \cite{c12}. The pion mass
dependence of the nucleon mass to fourth order is shown in Fig.~\ref{figmass4}.
We see that for pion masses below 400 MeV, the fourth order correction is
fairly small, for the physical value of $m_\pi$ these fourth order
corrections are so small that they can safely be neglected.
For larger values of $m_\pi$, the uncertainty due to the
variation in $e_1 (\lambda)$ becomes quickly very large. This clearly shows that it is
very dangerous to use a formula like Eq.~(\ref{HBmass4}) for too large pion
masses if one has no means to pin down the LEC $e_1$ with some accuracy.
Fortunately, from the analysis of pion--nucleon scattering to fourth order
we know  that $e_1$ should be of order 1~GeV$^{-3}$ and negative~\cite{c15}.
We note  that
there is some cancellation between the $m_\pi \ln m_\pi$ and the $m_\pi$ contributions
in $\delta M^{(4)}$ for negative values of $e_1$. If one e.g. changes the value of the
$c_3$ to  $-4\,$GeV$^{-1}$, the nucleon mass increases monotonically with increasing pion 
mass, as shown by the dotted line in Fig.~\ref{figmass4}, a trend consistent
with recent lattice data \cite{CPPACS}

\begin{figure}[h]
  \begin{center}
    \includegraphics*[width=0.45\textwidth]{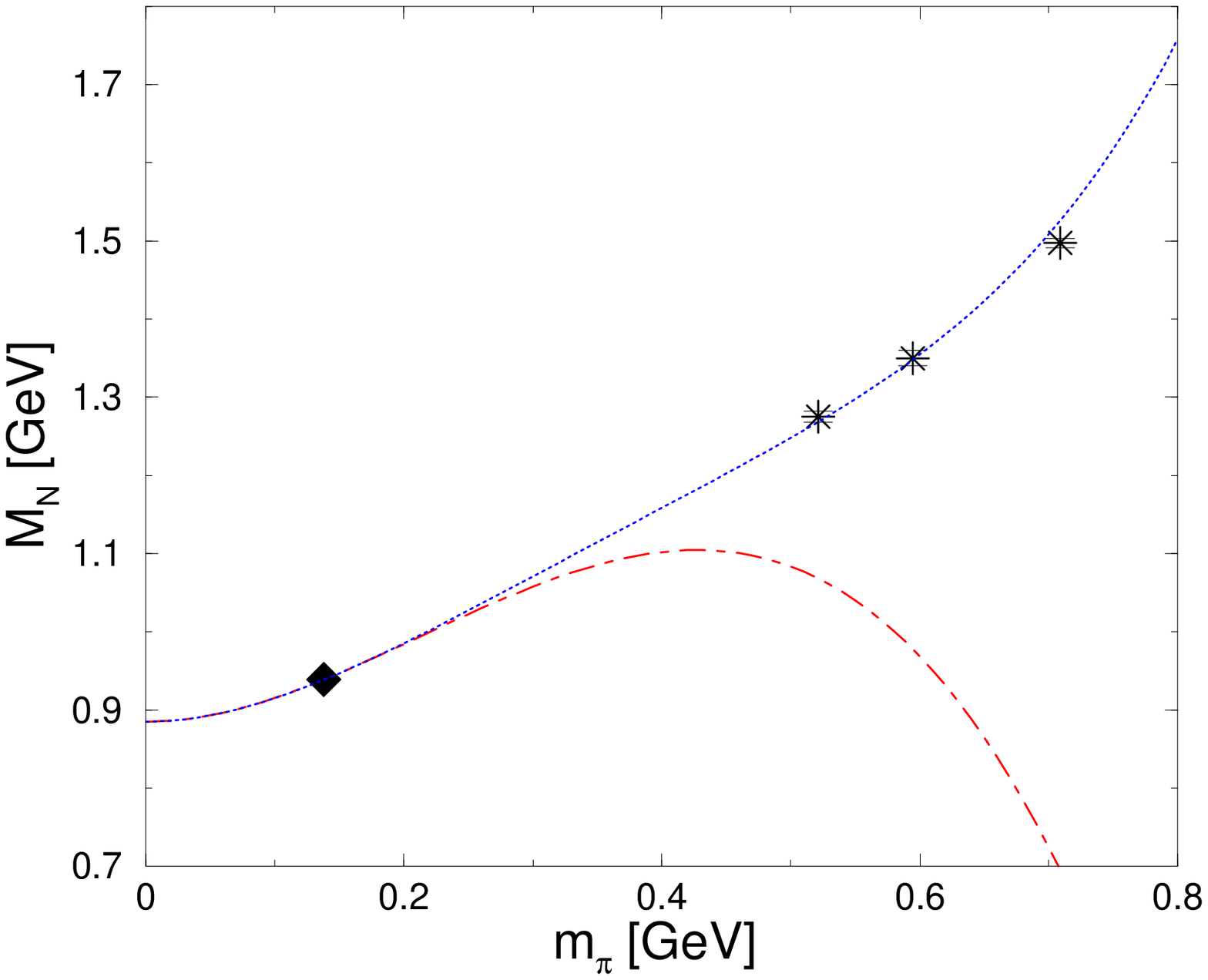}
    \includegraphics*[width=0.45\textwidth]{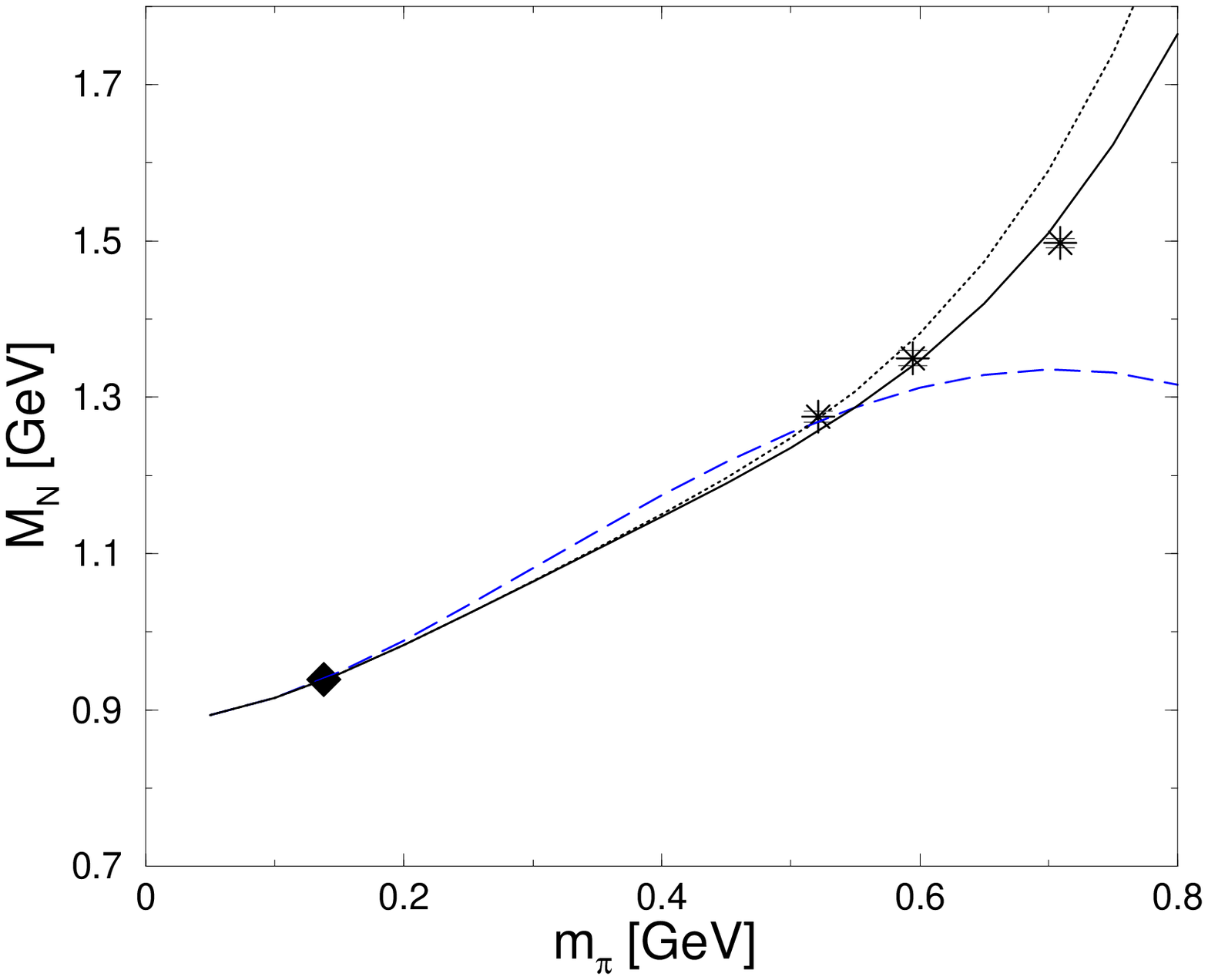}
    \caption{Chiral extrapolation of the nucleon mass. Left panel: best fit
     to the CP-PACS lattice results \protect\cite{CPPACS} employing dimensional
     regularization. The dot--dashed line refers to the parameter--free third
     order result whereas the dotted line gives the fit based on the fourth
     order representation for the values of the LECs given in the text.
     Right panel: Best fit in cutoff regularization. The dashed line gives 
     the improved third order result with one adjusted parameter. The solid (dotted)
     line refers to the fit based on the fourth order representation for the values of 
     the LECs given in the text with $\Lambda =1 \, (1.2)\,$GeV.
     }
\label{bestfit}
  \end{center}
\end{figure}

\medskip\noindent
In Fig.~\ref{bestfit} (left panel) we show the pion mass dependence of the
nucleon mass based on the dimensional regularization and allowing for variation
in the LEC $c_3$  to fit the lattice ``data'' from the CP-PACS collaboration
\cite{CPPACS}. We use $c_3 = -3.45\,\rm GeV^{-1}$, $e_1 = -1\,\rm GeV^{-3}$,
and all other parameters
and LECs as stated before. Note that for $g_A$ and $F_\pi$, we use the physical
values. One could also use the corresponding values in the chiral limit, which only
leads to a small change in the fit value for $c_3$, as we have convinced ourselves.
It is rather obvious that contrary to statements made in the literature, the fourth
order representation employing dimensional regularization can describe the quark
mass dependence of the lattice results without any need of unnatural values for any
of the  LECs~\footnote{The physics behind the seemingly large value of some of the
$c_i$ is well understood, see \cite{c12}.} or the
inclusion of higher order terms not generated in a symmetry--preserving one--loop
calculation. It should, however, be stressed again that for pion masses above 600~MeV
the fourth order corrections become unpleasantly large. 
In the right panel of  Fig.~\ref{bestfit}, we show the results employing
cutoff regularization (with the third order result including the improvement
term as discussed earlier). Again, the trend of the lattice results is nicely
reproduced setting $c_3 = -3.9\,\rm GeV^{-1}$ and $e_1 = -0.75\,\rm GeV^{-3}$,
which are modest changes to the values used in the case of dimensional regularization.
Furthermore, these values are within uncertainties consistent with extractions from
pion--nucleon scattering. Note that the corrections going from third to fourth order
are sizeably smaller than in case of dimensional regularization, which is mostly
due to the improvement term included in the third order result.
We point out again that in contrast to
statements found in the literature, these one--loop cutoff representations
are quite capable of describing the pion mass dependence of the nucleon mass
found in lattice gauge theory calculations.
The dependence on the actual choice of the cut--off $\Lambda$ is also shown in the
right panel of  Fig.~\ref{bestfit}, where the dotted line refers to the fourth order
result with $\Lambda =1 .2\,$GeV.
For the expression based on dimensional regularization to work above 600~MeV,
one has to include the sixth order term as in the improved cut--off scheme. 
We stress again that applying the one--loop expressions to pion masses above 600~MeV
is only done for illustrative purposes, for a realistic chiral extrapolation smaller
pion masses are mandatory. For a more detailed study of nucleon mass chiral extrapolations
based on such and similar regularizations, see \cite{extrapolation3}. 

\noindent
To further sharpen these statements, it is constructive to consider the
convergence of the chiral expansion for the nucleon mass as the pion mass
is increased. In Table~\ref{tab1} we have collected the pertinent results.
Consider first dimensional regularization. As it is well known and has been 
stressed before, for the physical value of the pion mass the convergence 
is excellent, and it is even fairly good for $m_\pi = 300\,$MeV.
Only for the large mass of 600~MeV, one has sizeable cancellations between the
second and third order pieces accompanied with a non--negligible fourth order
contribution. This is very reminiscent of the chiral expansion of the ground
state octet masses in terms of kaon and eta loop contributions. In cutoff 
regularization. the convergence properties are very visibly improved, so that
even at $m_\pi = 600\,$MeV the series is fairly well behaved.

\renewcommand{\arraystretch}{1.3}
 
\begin{table}[htb] 
\begin{center}
\begin{tabular}{||l||c||c|c||c|c||}
    \hline \hline
 & \multicolumn{1}{|c||}{} 
    & \multicolumn{2}{|c||} {Dim. Reg.}   &  \multicolumn{2}{|c||}
{Cutoff}    \\ 
    \hline  
$m_\pi$ [GeV] & ${\mathcal O}(q^2$)&
  ${\mathcal O}(q^3$)  & ${\mathcal O}(q^4)$  &  ${\mathcal O}(q^3$)  
  & ${\mathcal O}(q^4)$
  \\
\hline  \hline
$0.138$ &0.069& $-$0.015 & $-$0.0007 & $-$0.014 & $-$0.002  \\
 $0.300$ &0.32&  $-$0.151& 0.014 & $-$0.128 & $-$0.017   \\
 $0.600$  &1.30& $-$1.212 & 0.386 & $-$0.869& 0.034  \\
    \hline
    \hline  
\end{tabular}
\end{center}
\caption{Convergence of the chiral series for the nucleon mass employing
dimensional and cutoff regularization. The various contributions are given 
in GeV.}
\label{tab1}
\end{table}

\section{Criteria for a stable cutoff scheme}
\label{sec:test}

Based on the example discussed above, we can set up a general checklist 
for a consistent cutoff regularization procedure. It consists of the following steps:

\medskip\noindent
{\bf Step 1:} 
Choose a cutoff scheme and calculate all diagrams to the desired chiral order 
in that scheme.

\smallskip\noindent
{\bf Step 2:}
 Absorb all polynomial ($\sim \Lambda^n$) and logarithmic ($\sim \ln\Lambda$) 
divergences appearing to that order in 
counter terms. The finite (scale--dependent) parts of the corresponding 
LECs have to be kept as part of the full result, 
even when nothing is known about their size.

\smallskip\noindent
{\bf Step 3:}
Perform the chiral symmetry check: Expand the finite result in a series
in inverse powers of the cutoff  $\sim \Lambda^{-n}$ and 
check if the associated quark mass dependence is analytic, 
i.e. one must be able to show that all resulting terms can be matched onto an 
allowed contact interaction (counter term). 
If this is not the case, choose a different cutoff scheme or 
introduce extra  terms (see e.g. \cite{GJWL}) to correct for chiral 
symmetry violation.

\smallskip\noindent
{\bf Step 4:}
In case that the constraints from chiral symmetry are fulfilled, study the scale
dependence  of the full (unexpanded) finite result as a function of the cutoff scale.
Look for the onset of the plateau. If  the plateau is reached for values 
$\Lambda<\Lambda_\chi$ the calculation is consistent.

\smallskip\noindent
{\bf Step 5:}
If no plateau is found within the region of validity of the chiral
 effective field theory improvement terms {\em must} be added. The (negative of the)
 leading  term in the Taylor expansion in the cutoff scale discussed in step~3 
 provides the scale dependence for the local improvement term. The corresponding operator
 structure of the improvement term can also be read off from the Taylor expansion.
 As this term is analytic in the quark mass and allowed by chiral symmetry, 
the existence of this operator as part of a chiral Lagrangian is guaranteed.

\smallskip\noindent
{\bf Step 6:}
Add the improvement term to the unexpanded finite result. 
Note that this improvement term also carries a finite part of unknown
strength, as there is no 
free lunch in this world (or in other words: You cannot fool QCD into a reduced 
number of possible short distance operators, this only works in
models). Analyze  the remaining scale dependence: if the plateau now begins for cutoff values 
below the breakdown scale of the theory the addition of one improvement term was sufficient.
If that is not the case, go back to step 5 and continue the procedure until a satisfactory
result is obtained.

\medskip\noindent
This concludes our list to obtain stable prediction using cutoff regularization. 

\section{Summary and conclusions}
\label{sec:sum}

We summarize the main results of our work:
\begin{itemize}
\item[1)] CHPT results for the baryon structure {\em always} depend on an interplay
  between long and short distance physics. The relative strength between 
these two sectors  depends in general  on the regularization scale chosen. 
Only the sum of the two contributions  gives a scale--independent result. 
This statement holds independent of the regularization scheme 
(cutoff, dimensional  regularization, $\ldots$) chosen.
\item[2)] As soon as one goes beyond a leading order tree level analysis, 
the internal structure of 
hadrons is build up in CHPT. Two mechanism contribute
  to the internal structure effects: higher dimensional operators consistent
  with chiral symmetry and quantum fluctuations associated with Goldstone
  boson effects. This ``finite size'' of the nucleon emerges in chiral EFTs irrespective 
of the regularization scheme chosen.
\item[3)] Cutoff schemes are legitimate regularization procedures in CHPT 
(or other chiral EFTs),  provided that 
one checks carefully that terms generated by the employed scheme which 
are in conflict with chiral symmetry 
have all been removed. In section~\ref{sec:test} we have proposed test criteria which 
indicate whether the cutoff scheme chosen is appropriate for the question under study.
\item[4)] If a cutoff procedure fails the test criteria specified in 
section~\ref{sec:test}, 
improvement terms have to be added in the Lagrangian where 
the problem (i.e. large scale dependence in the region of validity 
of the theory) occurs. 
In section~\ref{sec:improved} we have shown how 
such an $O(\Lambda^{-1}) $ improvement term may look like 
for the simple example of the nucleon mass to leading one--loop order in HBCHPT. 
Given that such  improvement terms can be generated {\em within the CHPT framework}, 
we see no necessity to resort to phenomenologically inspired regularization 
procedure (e.g. inserting an exponential suppression factor into the chiral 
loop integrals ``by hand''). If the unimproved cutoff result has been properly 
corrected for any chiral symmetry breaking terms, then the addition of the 
improvement term calculated 
from this result also automatically fulfills all the constraints of chiral symmetry. 
In particular, it will not violate any strictures arising from (chiral) Ward identities, 
which is not guaranteed when adding phenomenologically inspired regulators 
to chiral loop integrals.
\item[5)] Comparisons between (scale--dependent) pion loop results obtained in CHPT 
(or any variant thereof)  
and phenomenological models are difficult. Qualitative comparisons to
  results obtained via dimensional regularization can only be obtained 
for ``small'' values of $\lambda\leq 600$ MeV, where $\lambda$ is the scale of 
dimensional regularization.
\item[6)] For the large majority of one-loop results in CHPT we see no 
need to give up dimensional 
regularization as the tool of choice, as the results in their interplay 
between short and long distance physics properly account for the internal 
(finite size) structure of the nucleon, e.g. see Eq.~(\ref{radius}). 
This statement holds independent of the masses of the Goldstone bosons
involved, e.g. also for  ``heavy'' pions in lattice simulations or 
octet Goldstone bosons. 
However, we note that cutoff approaches can be particularly useful in those
{\em special} cases where the leading one--loop result obtained in dimensional 
regularization appears to be scale-independent {\em and} at the same time the
mass of the involved Goldstone boson is large. We note that the improvement
procedures outline here can lead to a revived
analysis of SU(3) chiral dynamics, where for example the baryon mass 
calculation is exactly plagued by these problems.
\item[7)] We have analyzed the quark mass expansion of the nucleon mass
to second order in the quark masses in dimensional as well as in cutoff 
regularization. This representation can be used successfully to describe
lattice data obtained for pion masses between 500 and 750~MeV, although
for theoretical reasons such a  chiral extrapolation should at most be trusted
up to masses of about 600~MeV. Our results clearly rule out statements found in
the literature that one needs terms of very high orders to construct a
sensible chiral extrapolation function, furthermore, in our calculation chiral
symmetry is preserved at each step. Although these findings are promising, we
urge the lattice practitioners to generate results for pion masses below
500~MeV to really make contact to the chiral properties of QCD.
\end{itemize}
 
\bigskip

\section*{Acknowledgements}
One of the authors  (TRH) acknowledges lively discussions with G. Schierholz 
and the hospitality of the Forschergruppe ``Gitter und Hadronen
Physik'' 
at the University of Regensburg, where part of this work was completed. TRH also 
gratefully 
acknowledges the hospitality of and financial support from the ECT*  and the
Groupe de Physique Th\'eorique at Universit\'e Louis Pasteur de Strasbourg.

\appendix

\section{Third order calculation of the nucleon isovector anomalous magnetic moment}
\def\theequation{\Alph{section}.\arabic{equation}}
\setcounter{equation}{0}
\label{sec:mm}

\begin{figure}[ht]
  \begin{center}
    \includegraphics*[width=0.5\textwidth]{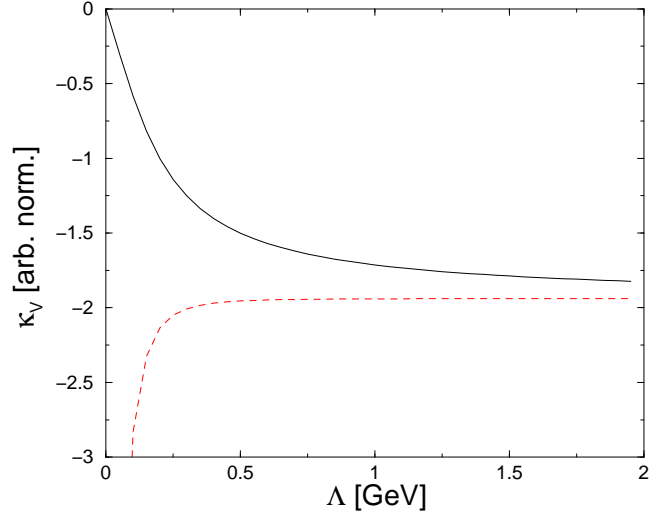}
    \caption{Cutoff dependence of the isovector nucleon anomalous magnetic moment 
             calculated at ${\mathcal O}(p^3)$ 
             in HBCHPT employing cutoff regularization (with an arbitrary normalization). 
             The result without (with)
             an improvement term as discussed in the text is shown by the solid (dashed)
             line.}
\label{figmmcut}
  \end{center}
\end{figure}

Here, we wish to discuss the isovector anomalous 
magnetic moment of the nucleon, because it is also
given to leading one--loop order by a simple formula but shows a somewhat different behaviour
than the nucleon mass. In dimensional regularization, one has 
\begin{equation}
\kappa_V= c_6 - \frac{g_A^2 M_N \, m_\pi }{4 \pi F_\pi^2}~,
\end{equation}
where the LEC $c_6$ can be adjusted so as to reproduce the physical value of the
isovector anomalous magnetic moment,  $\kappa_V = 3.706$. This leads to  $c_6 = 5.62$, which
seems unnaturally large but can be understood in terms of resonance saturation \cite{c12}.
Clearly, the third order result can only be used for pion masses below 400 MeV, because
for such a value the loop contribution completely cancels the leading dimension two
term $\sim c_6$. In fact, to third order the quark mass dependence of $\kappa_V$ is rather
trivial, the isovector magnetic moment simlpy decreases linearly with increasing $m_\pi$. 
As before, we now want to use a cutoff instead of dimensional regularization.
Using again a three--dimensional cutoff, we find
\begin{equation} \label{mmcut}
\kappa_V= c_6^{(r)} -{\frac{g_A^2 m_\pi M_N}{4 \pi F_\pi^2}} + 
{\frac{g_A^2  M_N}{2 \pi^2 F_\pi^2}} \left(m_\pi \arctan {
\frac{m_\pi}{\Lambda}}
+{\frac{1}{3}} {\frac{\Lambda m_\pi^2}{\Lambda^2 + m_\pi^2}}\right)
\end{equation}
with 
\begin{equation}
c_6^{(r)} = c_6+ {\frac{g_A^2  M_N}{3 \pi^2 F_\pi^2}}\Lambda~.
\end{equation}
We note that cutoff dependence is significantly different from the case of the
nucleon mass because of the appearance of terms linear in $m_\pi /\Lambda$.
This is shown in Fig.~\ref{figmmcut}, where even for the physical values of the
pion mass we do not find a plateau for cutoff values below the chiral symmetry breaking
scale (solid line, for simplicity we have set $c_6 = 0$). 
Needless to say that this cutoff dependence is even stronger for larger pion
masses.

\noindent
Expanding Eq.~(\ref{mmcut}) in inverse powers of $\Lambda$,  one gets:
\begin{equation}
\kappa_V=c_6^{(r)} -{\frac{g_A^2 m_\pi M_N}{4 \pi F_\pi^2}}
+{\frac{g_A^2  M_N}{3 \pi^2 F_\pi^2}} {\frac{m_\pi^2}{\Lambda}}
\left\{2 -{\frac{m_\pi^2}{\Lambda^2}} + \cdots \right\}~.
\end{equation}
From this expression, one can readily deduce the improvement term to be subtracted.
Again, the corresponding operator appears in the fourth order pion--nucleon Lagrangian,
it is the term $e_{105} \bar\psi \langle F_{\mu\nu}^+ \rangle \, \langle \chi_+
\rangle$ $\sigma^{\mu\nu}\, \psi\,$ \cite{FMMS}. Subtracting this contribution as before
and setting the finite part to zero for simplicity, one obtains the dashed curve in
Fig.~\ref{figmmcut}, which exhibits a nice plateau for the relevant values of the cutoff,
$2m_\pi \leq \Lambda \leq \Lambda_\chi$.



\begin{thebibliography}{99}
\bibitem{Weinberg} 
S.~Weinberg,
Physica A {\bf 96} (1979) 327.\vs
\bibitem{GL1} 
J.~Gasser and H.~Leutwyler,
Annals Phys.\  {\bf 158} (1984) 142.\vs
\bibitem{GL2} 
J.~Gasser and H.~Leutwyler,
Nucl.\ Phys.\ B {\bf 250} (1985) 465.\vs
\bibitem{Wein2}S.~Weinberg,
Phys.\ Rev.\  {\bf 166} (1968) 1568.\vs
\bibitem{CCWZ1} 
S.~R.~Coleman, J.~Wess and B.~Zumino,
Phys.\ Rev.\  {\bf 177} (1969) 2239.\vs
\bibitem{CCWZ2} 
C.~G.~Callan, S.~R.~Coleman, J.~Wess and B.~Zumino,
Phys.\ Rev.\  {\bf 177} (1969) 2247.\vs
\bibitem{Amherst1} 
J.~F.~Donoghue and B.~R.~Holstein,  Phys.\ Lett.\ B {\bf 436} (1998) 331.\vs
\bibitem{Amherst2} 
J.~F.~Donoghue, B.~R.~Holstein and B.~Borasoy,
Phys.\ Rev.\ D {\bf 59} (1999) 036002.\vs
\bibitem{Adelaide} 
R.~D.~Young, D.~B.~Leinweber and A.~W.~Thomas,
arXiv:hep-lat/0212031.\vs
\bibitem{Adelaide2}
D.~B.~Leinweber, A.~W.~Thomas and R.~D.~Young,
arXiv:hep-lat/0302020.\vs
\bibitem{GZ}
J.~Gasser and A.~Zepeda,
Nucl.\ Phys.\ B {\bf 174} (1980) 445.\vs
\bibitem{Juerg} J.~Gasser,
Annals Phys.\  {\bf 136} (1981) 62.\vs
\bibitem{EM}
D.~Espriu and J.~Matias,
Nucl.\ Phys.\ B {\bf 418} (1994) 494.\vs
\bibitem{HBCHPT}V.~Bernard, N.~Kaiser and U.-G.~Mei{\ss}ner,
Int.\ J.\ Mod.\ Phys.\ E {\bf 4} (1995) 193.\vs
\bibitem{SSE} T.~R.~Hemmert, B.~R.~Holstein and J.~Kambor,
J.\ Phys.\ G {\bf 24} (1998) 1831.\vs
\bibitem{ET} P.~J.~Ellis and H.~B.~Tang,
Phys.\ Rev.\ C {\bf 56} (1997) 3363.\vs
\bibitem{BL}
T.~Becher and H.~Leutwyler,
Eur.\ Phys.\ J.\ C {\bf 9} (1999) 643.\vs
\bibitem{BHM}
 V.~Bernard, T.~R.~Hemmert and U.-G.~Mei{\ss}ner,
Phys.\ Lett.\ B {\bf 565} (2002) 137.\vs
\bibitem{spin1} V.~Bernard, T.~R.~Hemmert and  U.-G. Mei{\ss}ner,
Phys.\ Lett.\ B {\bf 545} (2002) 105.\vs
\bibitem{spin2} V.~Bernard, T.~R.~Hemmert and  U.-G. Mei{\ss}ner,
Phys.\ Rev.\ D {\bf 67} (2003) 076008.\vs
\bibitem{BKKM} V.~Bernard, N.~Kaiser, J.~Kambor and U.-G. Mei{\ss}ner,
Nucl.\ Phys.\ B {\bf 388} (1992) 315.\vs
\bibitem{MMD} P.~Mergell,   U.-G. Mei{\ss}ner and D.~Drechsel,
Nucl.\ Phys.\ A {\bf 596} (1996) 367.\vs
\bibitem{HMD} 
H.~W.~Hammer, U.-G. Mei{\ss}ner and D.~Drechsel,
Phys.\ Lett.\ B {\bf 385} (1996) 343.\vs
\bibitem{BFHM} V. Bernard, H.W. Fearing, T.R. Hemmert and U.-G. Mei{\ss}ner, Nucl. Phys.
Nucl.\ Phys.\ A {\bf 635} (1998) 121
[Erratum-ibid.\ A {\bf 642} 563].\vs
\bibitem{KM}B.~Kubis and U.-G.~Mei{\ss}ner,
Nucl.\ Phys.\ A {\bf 679} (2001) 698.\vs
\bibitem{masses1} 
J.~Gasser, H.~Leutwyler and M.~E.~Sainio,
Phys.\ Lett.\ B {\bf 253} (1991) 260.\vs
%
\bibitem{masses2} 
E.~Jenkins and A.~V.~Manohar,
Phys.\ Lett.\ B {\bf 281} (1992) 336.\vs
\bibitem{masses3} 
B.~Borasoy and U.-G.~Mei{\ss}ner,
Annals Phys.\  {\bf 254} (1997) 192.\vs
\bibitem{masses4} 
B.~Borasoy,
Eur.\ Phys.\ J.\ C {\bf 8} (1999) 121.\vs
\bibitem{Gerrit} G. Schierholz et al. (QCDSF collaboration), in preparation.\vs
\bibitem{c11}
V.~Bernard, N.~Kaiser and U.-G. Mei{\ss}ner,
Nucl.\ Phys.\ B {\bf 457} (1995) 147.\vs
\bibitem{c12}
V.~Bernard, N.~Kaiser and U.-G. Mei{\ss}ner, 
Nucl.\ Phys.\ A {\bf 615} (1997) 483.\vs
\bibitem{c13}
M.~Moj\v zi\v s,
Eur.\ Phys.\ J.\ C {\bf 2} (1998) 181.\vs
\bibitem{c14}
N.~Fettes, U.-G. Mei{\ss}ner and S.~Steininger,
Nucl.\ Phys.\ A {\bf 640} (1998) 199.\vs
\bibitem{c15}
N.~Fettes and U.-G. Mei{\ss}ner,
Nucl.\ Phys.\ A {\bf 676} (2000) 311.\vs
\bibitem{c16}
T.~Becher and H.~Leutwyler,
JHEP {\bf 0106} (2001) 017.\vs
\bibitem{c17}
K.~Torikoshi and P.~J.~Ellis,
Phys.\ Rev.\ C {\bf 67} (2003) 015208.\vs
\bibitem{CPPACS}
A.~Ali Khan {\it et al.}  [CP-PACS Collaboration],
Phys.\ Rev.\ D {\bf 65} (2002) 054505
[Erratum-ibid.\ D {\bf 67} (2003) 059901].\vs
\bibitem{extrapolation1}
T.~R.~Hemmert and W.~Weise,
Eur.\ Phys.\ J.\ A {\bf 15} (2002) 487.\vs
\bibitem{extrapolation2}
T.~R.~Hemmert, M.~Procura and W.~Weise,
arXiv:hep-lat/0303002.\vs
\bibitem{EGMreg}E.~Epelbaum, W.~Gl\"ockle and U.-G.~Mei{\ss}ner,
arXiv:nucl-th/0304037.\vs
\bibitem{FMMS}
N.~Fettes, U.-G.~Mei{\ss}ner, M.~Moj\v zi\v s and S.~Steininger,
Annals Phys.\  {\bf 283} (2000) 273
[Erratum-ibid.\  {\bf 288} (2001) 249].\vs
\bibitem{hwh1}
P.~F.~Bedaque, H.~W.~Hammer and U.~van Kolck,
Phys.\ Rev.\ Lett.\  {\bf 82} (1999) 463.\vs
\bibitem{hwh2}
P.~F.~Bedaque, H.~W.~Hammer and U.~van Kolck,
Nucl.\ Phys.\ A {\bf 646} (1999) 444.\vs
\bibitem{SFM}
S.~Steininger, U.-G.~Mei{\ss}ner and N.~Fettes,
JHEP {\bf 9809} (1998) 008.\vs
\bibitem{MGB}
J.~A.~McGovern and M.~C.~Birse,
Phys.\ Lett.\ B {\bf 446} (1999) 300.\vs
\bibitem{extrapolation3}
M.~Procura, T.~R.~Hemmert  and W.~Weise, in preparation.\vs
\bibitem{GJWL}
I.~S.~Gerstein, R.~Jackiw, S.~Weinberg and B.~W.~Lee,
Phys.\ Rev.\ D {\bf 3} (1971) 2486.\vs
\end{thebibliography}
\end{document}